\documentclass[aps,prd,twocolumn,nofootinbib]{revtex4-1}
\usepackage{epsfig}
\usepackage[colorlinks,linkcolor=blue,anchorcolor=blue,citecolor=blue,urlcolor=blue,breaklinks=true]{hyperref}
\usepackage{graphics}
\usepackage{slashed}
\usepackage{color}
\usepackage{amsmath}
\usepackage{bm}
\usepackage{setspace}
\emergencystretch=\maxdimen
\begin{document}
\author{Ya-Peng Zhao$^{1}$}\email{zhaoyapeng2013@hotmail.com}
\address{$^{1}$ Collage of Physics and Electrical Engineering, Anyang Normal University, Anyang, 455000, China}
\title{Thermodynamic properties and transport coefficients of QCD matter within the non-extensive Polyakov-Nambu-Jona-Lasinio model}

\begin{abstract}
   We present a non-extensive version of the Polyakov-Nambu-Jona-Lasinio model which is based on the non-extentive statistical mechanics. This new statistics is characterized by a dimensionless non-extensivity parameter $q$ that accounts for all possible effects violating the assumptions of the Boltzmann-Gibbs statistics (when $q\rightarrow1$, it returns to the Boltzmann-Gibbs case). Using this q-Polyakov-Nambu-Jona-Lasinio model and including two different Polyakov-loop potentials, we discussed the influence of the parameter $q$ on chiral and deconfinement phase transition, various thermodynamic quantities and transport coefficients at finite temperature and zero quark chemical potential.
   We found that the Stefan-Boltzmann limit is actually related to the choice of statistics. For example, in the Tsallis statistics, the thermodynamic quantities $\frac{\epsilon}{T^{4}}$, $\frac{p}{T^{4}}$ and $\frac{s}{T^{3}}$ all increase with $q$, exceed their usual Stefan-Boltzmann limits and tend to a new $q$-related Tsallis limit at temperature high enough. Interestingly, however, due to a surprising cancellation, the high temperature limit of $c_{s}^{2}$ is still its SB limit $1/3$. In addition, we found some similarities between the non-extensive effect and the finite-size effect. For example, as $q$ increases (size decreases), the criticality of $\frac{c_{v}}{T^{3}}$ and $c_{s}^{2}$ gradually disappears.
   Besides, in order to better study the non-extensive effect, we defined a new susceptibility and calculated the response of thermodynamic quantities and transport coefficients to $q$. And found that their response patterns are different.
\bigskip

\bigskip

\noindent PACS Number(s): 12.38.Mh, 12.39.-x, 25.75.Nq, 12.38.Aw

\end{abstract}


\maketitle

\section{INTRODUCTION}
Among all standard studies of the QCD matter, a statistical approach often used is Boltzmann-Gibbs (BG) statistics. However, strictly speaking, this approach is correct only when the corresponding heat bath is homogeneous and infinite. Obviously, in reality, this condition cannot always be met. Especially in the relativistic heavy-ion collisions, in which the quark-gluon plasma (QGP) produced experiences strong intrinsic fluctuations and long-range correlations. The size of QGP is small enough and it evolves rapidly. Therefore, this system is far from being uniform and no global equilibrium is established. As a result, some quantities become non-extensive and develop power-law tailed rather than exponential distributions. In such cases the application of the usual BG statistics is questionable.

Thus, a non-extensive statistics that extended BG statistics was first proposed by Tsallis~\cite{Tsallis1988}. The most typical feature of Tsallis statistics is that it replaces the usual exponential factors by their q-exponential equivalents~\cite{PhysRevC.77.044903,Eur.Phys.J.A2016,shen2017chiral},
\begin{eqnarray}\label{Tsallis}
P_{BG}(E)=exp(-\frac{E}{T})\longrightarrow P_{q}(E)=exp_{q}(-\frac{E}{T}),
\end{eqnarray}
where
\begin{eqnarray}
exp_{q}(x)=[1+(1-q)x]^{\frac{1}{1-q}},
\end{eqnarray}
correspondingly, its inverse function is
\begin{eqnarray}
ln_{q}(x)=\frac{x^{1-q}-1}{1-q}.
\end{eqnarray}
The non-extensivity parameter $q$ represents all possible factors that do not satisfy the BG statistical assumptions. When $q\rightarrow1$, $exp_{q}(x)\rightarrow exp(x)$, $ln_{q}(x)\rightarrow ln(x)$ and Tsallis statistics returns to BG statistics.

In high-energy physics, using Tsallis statistics to describe the transverse momentum distributions is now a standard practice~\cite{BEDIAGA2000156,Wilk2012,LI2013352,PhysRevD.91.054025,De2014,Bhattacharyya2016}. It is excellent in meeting the experimental data, as pointed out by the PHENIX~\cite{PhysRevC.83.064903} and STAR~\cite{PhysRevC.75.064901} Collaborations at RHIC and by the ALICE~\cite{Aamodt2011}, ATLAS~\cite{aad2011charged} and CMS~\cite{Khachatryan2010} Collaborations at the LHC.
In addition, more and more physical branches, even biology, economics are described by Tsallis distribution. A general overview on Tsallis' statistics and its diverse applications can be found in Ref.~\cite{tsallis2009introduction}. Finally, it should be noted that Ref.~\cite{PhysRevC.97.044913} also raises doubts about the application of Tsallis statistics to relativistic heavy-ion physics.

Studying the thermodynamic properties and transport coefficients of the QCD matter has always been a matter of great interest to people. At high baryon density and low temperature, they are relevant to the study of compact stars~\cite{Buballa_2014,Baym_2018,PhysRevD.99.043001,PhysRevD.81.123016}. For example, the equation of state can be combined with the Tolman-Oppenheimer-Volkoff (TOV) equation to study the mass-radius relationship, the internal structure of compact stars and further to study the tidal Love number $k_{2}$ and tidal deformability $\Lambda$.
At high temperature and low baryon density, they are relevant to the QGP produced in relativistic heavy-ion collisions.
Especially, a low value of the shear viscosity to entropy ($\eta/s$) is needed to explain the elliptic flow data~\cite{PhysRevLett.99.172301}, which means the QGP is actually a strongly-coupled medium.
In addition, studies on thermodynamic quantities and transport coefficients may also help to reveal QCD phase transitions or a rapid crossover~\cite{PhysRevLett.103.172302,PhysRevC.74.014901,KARSCH2008217,Zhang_2018}.

In this paper, the question we are concerned with is that when a strongly interacting system is described by Tsallis statistics, what is the difference between the thermodynamic quantities and transport coefficients and that of BG statistics. For this purpose, we generalize the PNJL model to its non-extensive version. Compared with NJL model, this model has proven to be more successful in reproducing lattice data concerning QCD thermodynamics~\cite{PhysRevD.73.014019}. Besides, other models such as the linear sigma model and NJL model have also been generalized to its non-extensive version to study the thermodynamic quantities of the QCD matter and its phase diagram~\cite{Eur.Phys.J.A2016,shen2017chiral}.

This paper is organized as follows: In Sec. II, we introduce the non-extensive version of the PNJL model and discuss the q-dependence of the chiral and deconfinement phase transition at finite temperature and zero quark chemical potential.
In Sec. III we analyze in detail the influence of the parameter $q$ on the thermodynamic quantities and transport coefficients.
Finally, we give a brief summary of our work in Sec. IV.
\section{non-extensive pnjl model:q-pnjl}\label{ok}
Before introducing the q-PNJL model, let's make a basic introduction to the PNJL model.
The Lagrangian of the two-flavor and three-color PNJL model reads~\cite{PhysRevD.73.014019}
\begin{eqnarray}
\mathcal{L}_{PNJL} &=& \bar{\Psi}(i\gamma_{\mu}D^{\mu}-\hat{m})\Psi +G\,[(\bar{\Psi}\Psi)^2+(\bar{\Psi}i\gamma_5\vec{\tau}\Psi)^2]\nonumber\\ &&-\mathcal{U}(\Phi,\bar{\Phi};T),
\label{effective}
\end{eqnarray}
where $\Psi=(u,d)$ and $\hat{m}=diag(m_{u},m_{d})$ with $m_{\mu}=m_{d}=m$ stands for the current quark mass matrix. $\tau^{a}(a=1,2,3)$ are Pauli matrices acting in flavor space and $G$ is the effective coupling strength of four point interaction of quark fields. The effective Polyakov-loop potential $\mathcal{U}(\Phi,\bar{\Phi};T)$ accounts for the self-interactions of the gauge field in which the normalized color-traced Polyakov-loop expectation value $\Phi$ and its Hermitian conjugation $\bar{\Phi}$ are defined as
\begin{eqnarray}
\Phi=\frac{\langle Tr_{c}L \rangle}{N_{c}},     \bar{\Phi}=\frac{\langle Tr_{c}L^{\dagger} \rangle}{N_{c}},
\end{eqnarray}
where the Polyakov line is defined as
\begin{eqnarray}
L(\vec{x})=P\exp(i\int_{0}^{\beta}A_{4}(\vec{x},\tau)d\tau),
\end{eqnarray}
 and $A_{4}=iA_{0}$ is the temporal component of Euclidian gauge field $(\vec{A},A_{4})$, $\beta=\frac{1}{T}$, and $P$ denotes the path ordering.
The covariant derivative is determined as
\begin{eqnarray}
D_{\mu}&=&\partial_{\mu}-iA_{\mu},\nonumber \\
A_{\mu}&=&\delta_{\mu}^{0}A_{0},
\end{eqnarray}
here $A_{\mu}=gA_{\mu}^{a}\frac{\lambda^{a}}{2}$ and $g$ is the $SU(3)_{c}$ gauge coupling. The $\lambda^{a}$ stands for the Gell-Mann matrices with $\lambda^{0}=\sqrt{\frac{2}{3}}1$.

Under the mean-field approximation, the thermodynamic potential density function is
\begin{eqnarray}
\Omega(\mu,T,M,\Phi,\bar{\Phi})&=&\mathcal{U}(\Phi,\bar{\Phi};T)+\frac{(M-m)^{2}}{4G}\\
&-&2N_{c}N_{f}\int_{0}^{\Lambda}\frac{{\rm d}^3\vec{p}}{(2\pi)^3}E_{p}\nonumber\\
&-&2N_{f}T\int_{0}^{\infty}\frac{{\rm d}^3\vec{p}}{(2\pi)^3}(lnF^{+}+lnF^{-}),\nonumber
\end{eqnarray}
where $M$ means the dynamical quark mass. It relates to the quark chiral condensate $\sigma=\langle\bar{\Psi}\Psi\rangle$ as follows
\begin{eqnarray}
M=m-2G\sigma,
\end{eqnarray}
and
\begin{eqnarray}
F^{+}&=&1+3(\Phi+\bar{\Phi}e^{-\frac{E_{p}-\mu}{T}})e^{-\frac{E_{p}-\mu}{T}}+e^{-3\frac{E_{p}-\mu}{T}},\nonumber\\
F^{-}&=&1+3(\bar{\Phi}+\Phi e^{-\frac{E_{p}+\mu}{T}})e^{-\frac{E_{p}+\mu}{T}}+e^{-3\frac{E_{p}+\mu}{T}},
\end{eqnarray}
in which $E_{p}=\sqrt{p^{2}+M^{2}}$ is the single quasi-particle energy. In the above integrals, following Refs.~\cite{sym2031338,PhysRevD.73.014019,PhysRevC.79.055208,FUKUSHIMA2004277}, the vacuum integral has a cut-off $\Lambda$ whereas the medium dependent integrals have been extended to infinity.

Finally, the solutions of the mean field equations are obtained by minimizing the thermodynamic potential function $\Omega$ with respect to $M$, $\Phi$ and $\bar{\Phi}$, that is
\begin{eqnarray}\label{gap}
\frac{\partial\Omega}{\partial M}=\frac{\partial\Omega}{\partial\Phi}=\frac{\partial\Omega}{\partial\bar{\Phi}}=0,
\end{eqnarray}
at vanishing chemical potential, $\bar{\Phi}=\Phi$.
\subsection{Polyakov-loop potentials}
The functional form of the effective Polyakov-loop potential $\mathcal{U}$ that can be constructed from the center symmetry of the pure gauge sector is not unique. The required parameters are based on the pure gauge lattice data. Next two effective Polyakov-loop potentials are introduced.

The polynomial effective Polyakov-loop potential is~\cite{PhysRevD.62.111501,PhysRevC.66.034903,PhysRevD.73.014019}
\begin{eqnarray}
\frac{\mathcal{U_{P}}}{T^{4}}=-\frac{b_{2}(T)}{2}\bar{\Phi}\Phi-\frac{b_{3}}{6}(\Phi^{3}+\bar{\Phi}^{3})+\frac{b_{4}}{4}(\bar{\Phi}\Phi)^{2},
\end{eqnarray}
with a temperature-dependent coefficient
\begin{eqnarray}
b_{2}(T)=a_{0}+a_{1}(\frac{T_{0}}{T})+a_{2}(\frac{T_{0}}{T})^{2}+a_{3}(\frac{T_{0}}{T})^{3},
\end{eqnarray}
and the corresponding parameters are given in Table~\ref{tb1}.
\begin{center}
\begin{table}
\caption{Parameter set used in our work.}\label{tb1}
\begin{tabular}{p{1.2cm} p{1.2cm} p{1.2cm} p{1.2cm} p{1.2cm} p{1.2cm}}
\hline\hline
$a_0$&$a_1$&$a_2$&$a_3$&$b_3$&$b_4$\\
\hline
6.75&-1.95&2.625&-7.44&0.75&7.5\\
\hline\hline
\end{tabular}
\end{table}
\end{center}
In a pure gauge sector, $T_{0}=270\ \mathrm{MeV}$. However, in the presence of dynamical quarks, the critical temperature $T_{0}$ will have an $N_{f}$ dependence $T_{0}(N_{f})$. For massless flavors, $T_{0}(2)=208\ \mathrm{MeV}$ with an uncertainty of about $30\ \mathrm{MeV}$. If we consider that quark has mass, the critical temperature will be lower. Here, we let $T_{0}(2)=192\ \mathrm{MeV}$  follows Ref.~\cite{PhysRevD.76.074023}.
Besides, it should be noted that with this Polynomial potential $\mathcal{U_{P}}$, the Polyakov-loop expectation value $\Phi$ (at $\mu=0$, $\Phi=\bar{\Phi}$) will be greater than one at a temperature of a few hundred $\mathrm{MeV}$ and when $T\rightarrow\infty$, $\Phi\simeq1.11$.

The Logarithmic effective Polyakov-loop potential is~\cite{PhysRevD.75.034007}
\begin{eqnarray}
\frac{\mathcal{U_{L}}}{T^{4}}&=&-\frac{a(T)}{2}\Phi\bar{\Phi}+b(T)ln[1-6\Phi\bar{\Phi}-3(\Phi\bar{\Phi})^{2}\nonumber\\
&&+4(\Phi^{3}+\bar{\Phi}^{3})],
\end{eqnarray}
with the temperature-dependent coefficients
\begin{eqnarray}
a(T)=a_{0}+a_{1}(\frac{T_{0}}{T})+a_{2}(\frac{T_{0}}{T})^{2},
\end{eqnarray}
and
\begin{eqnarray}
b_{T}=b_{3}(\frac{T_{0}}{T})^{3},
\end{eqnarray}
the corresponding parameters are given in Table~\ref{tb2}. Here, the logarithmic form constrains $\Phi,\bar{\Phi}\leq1$.
\begin{center}
\begin{table}
\caption{Parameter set used in our work.}\label{tb2}
\begin{tabular}{p{1.8cm} p{1.8cm} p{1.8cm} p{1.8cm}}
\hline\hline
$a_0$&$a_1$&$a_2$&$b_3$\\
\hline
3.51&-2.47&15.2&-1.75\\
\hline\hline
\end{tabular}
\end{table}
\end{center}

\begin{center}
\begin{table}
\caption{Parameter set used in our work.}\label{tbnjl}
\begin{tabular}{p{2.4cm} p{2.4cm} p{2.4cm}}
\hline\hline
$\Lambda(\mathrm{MeV})$&$G(\mathrm{MeV^{-2}})$&$m(\mathrm{MeV})$\\
\hline
651&$5.04\times10^{-6}$&5.5\\
\hline\hline
\end{tabular}
\end{table}
\end{center}

Besides, the parameters for the NJL model part of the effective Lagrangian $\mathcal{L}_{PNJL}$ are summarized in Table~\ref{tbnjl}. The resulting physical quantities are $f_{\pi}=92.3\ \mathrm{MeV}$, $m_{\pi}=139.3\ \mathrm{MeV}$ and $-\langle\bar{\Psi}_{u}\Psi_{u}\rangle^{\frac{1}{3}}=251\ \mathrm{MeV}$~\cite{PhysRevD.73.014019}.
\subsection{q-PNJL model}
In short, when we use Tsallis statistics instead of BG statistics to describe a system, it means we need to do the replacement as shown in Eq.~(\ref{Tsallis}).
In this first case study we shall however take up the following two simplifications:

(i) in the present treatment non-extensive effects are not considered in the pure Yang-Mills sector. As a consequence, the Polyakov-loop potential remains unchanged and feels non-extensive effects implicitly only through the saddle point equations.

(ii) we shall not use any modifications to the usual PNJL model parameters due to non-extensive effects. We treat $q$ as a thermodynamic variable in the same footing as $T$ and $\mu$. Similarly, in the study of finite-size effects, they also treat volume $V$ as a thermodynamic variable in the same footing as $T$ and $\mu$. Fitting the parameters at $T=0$, $\mu=0$ and $V=\infty$ and then studying the finite-size effects at finite temperature or/and quark chemical potential~\cite{PhysRevD.87.054009,Grunfeld2018}. In fact, this is all based on the ansatz that the parameters determined at zero temperature and zero quark chemical potential can be used to study the finite temperature and finite quark chemical potential. Of course, it is also pointed out in the Refs.~\cite{PhysRevD.82.076003,Cui2014} that the parameter of the coupling constant $G$ should depend on the order parameter $\Phi$ or $\langle\bar{\Psi}\Psi\rangle$ and then implicitly on the temperature and the quark chemical potential. But here, we do not consider this situation.

Thus, within the q-PNJL model, the thermodynamic potential density function becomes
\begin{eqnarray}\label{omega00}
\Omega_{q}(\mu,T,M,\Phi,\bar{\Phi})&=&\mathcal{U}(\Phi,\bar{\Phi};T)+\frac{(M-m)^{2}}{4G}\\
&-&2N_{c}N_{f}\int_{0}^{\Lambda}\frac{{\rm d}^3\vec{p}}{(2\pi)^3}E_{p}\nonumber\\
&-&2N_{f}T\int_{0}^{\infty}\frac{{\rm d}^3\vec{p}}{(2\pi)^3}(ln_{q}F_{q}^{+}+ln_{q}F_{q}^{-}),\nonumber
\end{eqnarray}
where
\begin{eqnarray}
F_{q}^{+}=\nonumber\\
1+3(\Phi+\bar{\Phi}e_{q}(-\frac{E_{p}-\mu}{T}))e_{q}(-\frac{E_{p}-\mu}{T})+e_{q}(\frac{-3(E_{p}-\mu)}{T}),\nonumber\\
F_{q}^{-}=\nonumber\\
1+3(\bar{\Phi}+\Phi e_{q}(-\frac{E_{p}+\mu}{T}))e_{q}(-\frac{E_{p}+\mu}{T})+e_{q}(\frac{-3(E_{p}+\mu)}{T}).\nonumber\\
\end{eqnarray}

In order to ensure that $e_{q}(x)$ is always a non-negative real function, the following constraint must be met.
\begin{eqnarray}\label{Tsa}
[1+(1-q)x]\geq0.
\end{eqnarray}
And in this paper, as a first step, we consider only $q>1$. This is because on the one hand the typical value of the non-extensivity parameter $q$ for high energy collisions is found to be $1 \leq q \leq 1.2$~\cite{CLEYMANS2013351,LI2013352,PhysRevD.91.054025,Azmi_2014}. On the other hand in the case of $q>1$, $q-1$ is a measure of intrinsic fluctuations of the temperature in the system considered~\cite{PhysRevLett.94.132302,PhysRevLett.84.2770}, whereas $q<1$, the interpretation of $q-1$ is inconsistent~\cite{Kodama2009,GARCIAMORALES2006161}. So, in the case of zero quark chemical potential and finite temperature, the condition Eq.~(\ref{Tsa}) is naturally satisfied. If not, we can use the following Tsallis cut-off prescription for $q>1$
\begin{eqnarray}
e_{q}(x)=0, \quad \text{for} \quad [1+(1-q)x]<0,
\end{eqnarray}
or without Tsallis cut-off prescription
\begin{equation}
e_{q}(x)=
\begin{cases}
[1+(1-q)x]^{\frac{1}{1-q}}, \quad \text{for} \quad x\leq0,\\
[1+(q-1)x]^{\frac{1}{q-1}}, \quad \text{for} \quad x>0.
\end{cases}
\end{equation}
It should be pointed out that it is not clear so far under which circumstances Tsallis cut-off or without Tsallis cut-off should be used. A more detailed discussion can be found in Ref.~\cite{Eur.Phys.J.A2016}.

Besides, it is important to realize that for $T\rightarrow0$ one always gets $\Omega_{q}\rightarrow\Omega$ as long as $q>1$. This means that we can expect any non-extensive signature only for high enough temperatures.

For studying the phase transition within the q-PNJL model at zero quark chemical potential and finite temperature, according to Eq.~(\ref{gap}), the coupled non-linear equations for the $M$ and $\Phi$ can then be obtained as follows
\begin{widetext}
\begin{eqnarray}
M&=&m+4GN_{c}N_{f}\int\frac{{\rm d}^3\vec{p}}{(2\pi)^3}\frac{M}{E_{p}}[1-n_{q}-\bar{n}_{q}],\\
0&=&\frac{\partial\mathcal{U}}{\partial\Phi}-6N_{f}T\int_{0}^{\infty}\frac{{\rm d}^3\vec{p}}{(2\pi)^3}\{\frac{(1+e_{q}(\frac{-(E_{p}-\mu)}{T}))e_{q}(\frac{-(E_{p}-\mu)}{T})}{[1+3\Phi(1+e_{q}(\frac{-(E_{p}-\mu)}{T}))e_{q}(\frac{-(E_{p}-\mu)}{T})+e_{q}(\frac{-3(E_{p}-\mu)}{T})]^{q}}\nonumber\\
&&+\frac{(1+e_{q}(\frac{-(E_{p}+\mu)}{T}))e_{q}(\frac{-(E_{p}+\mu)}{T})}{[1+3\Phi(1+e_{q}(\frac{-(E_{p}+\mu)}{T}))e_{q}(\frac{-(E_{p}+\mu)}{T})+e_{q}(\frac{-3(E_{p}+\mu)}{T})]^{q}}\},
\end{eqnarray}
\end{widetext}
where the q-version of the Fermi-Dirac distribution is
\begin{eqnarray}\label{nq}
n_{q}(T,\mu)=\nonumber\\
\frac{e_{q}^{q}(\frac{-3(E_{p}-\mu)}{T})+\Phi(1+2e_{q}(\frac{-(E_{p}-\mu)}{T}))e_{q}^{q}(\frac{-(E_{p}-\mu)}{T})}{[1+3\Phi(1+e_{q}(\frac{-(E_{p}-\mu)}{T}))e_{q}(\frac{-(E_{p}-\mu)}{T})+e_{q}(\frac{-3(E_{p}-\mu)}{T})]^{q}},\nonumber\\
\end{eqnarray}
and
\begin{eqnarray}\label{nbarq}
\bar{n}_{q}(T,\mu)=\nonumber\\
\frac{e_{q}^{q}(\frac{-3(E_{p}+\mu)}{T})+\Phi(1+2e_{q}(\frac{-(E_{p}+\mu)}{T}))e_{q}^{q}(\frac{-(E_{p}+\mu)}{T})}{[1+3\Phi(1+e_{q}(\frac{-(E_{p}+\mu)}{T}))e_{q}(\frac{-(E_{p}+\mu)}{T})+e_{q}(\frac{-3(E_{p}+\mu)}{T})]^{q}},\nonumber\\
\end{eqnarray}
when $q\rightarrow1$, they return to the distribution function of the usual PNJL model.
\subsection{Finite-temperature QCD transition within the q-PNJL model}
\begin{center}
\begin{table}
\caption{The $q$ dependence of $T_{c}$ at effective Polyakov-loop potential $\mathcal{U_{P}}$.}\label{tb3}
\begin{tabular}{p{1.8cm} p{1.8cm} p{1.8cm} p{1.8cm}}
\hline\hline
$ $&$q=1$&$q=1.05$&$q=1.1$\\
\hline
$T_{\chi}(\mathrm{MeV})$&196&188&180\\
\hline
$T_{d}(\mathrm{MeV})$&170&164&154\\
\hline\hline
\end{tabular}
\end{table}
\end{center}
\begin{center}
\begin{table}
\caption{The $q$ dependence of $T_{c}$ at effective Polyakov-loop potential $\mathcal{U_{L}}$.}\label{tb4}
\begin{tabular}{p{1.8cm} p{1.8cm} p{1.8cm} p{1.8cm}}
\hline\hline
$ $&$q=1$&$q=1.05$&$q=1.1$\\
\hline
$T_{\chi}(\mathrm{MeV})$&200&194&188\\
\hline
$T_{d}(\mathrm{MeV})$&160&152&142\\
\hline\hline
\end{tabular}
\end{table}
\end{center}
\begin{figure}
\includegraphics[width=0.47\textwidth]{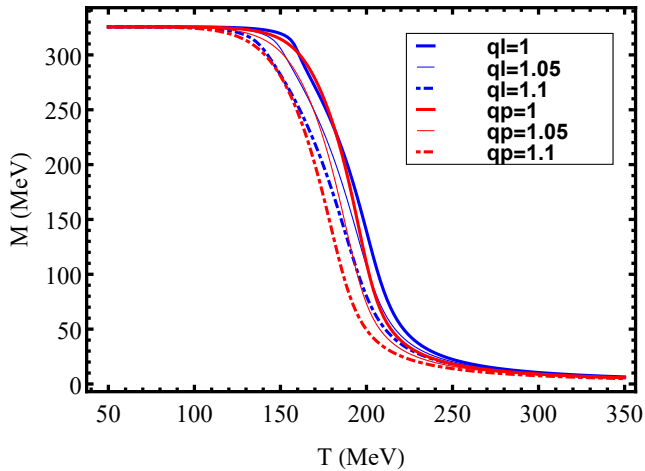}
\caption{Constituent quark mass $M$ as a function of $T$ at $\mu=0$ for two different potentials $\mathcal{U}$ and three parameters $q$. Where $\mathrm{l}$, and $\mathrm{p}$ represent logarithmic and polynomial Polyakov-loop potential, respectively.}
\label{Fig:M}
\end{figure}
\begin{figure}
\includegraphics[width=0.47\textwidth]{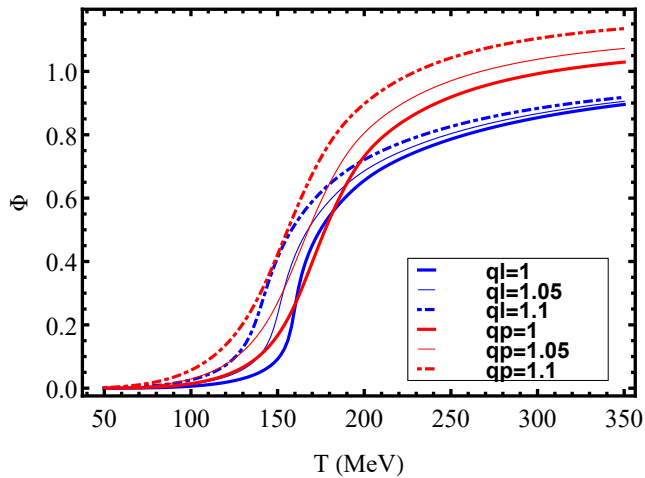}
\caption{Polyakov-loop expectation value $\Phi$ as a function of $T$ at $\mu=0$ for two different Polyakov-loop potentials $\mathcal{U}$ and three parameters $q$.}
\label{Fig:phi}
\end{figure}

The coupled non-linear equations can be numerically solved by iteration. In Figs.~\ref{Fig:M},~\ref{Fig:phi} we plot $M$ and $\Phi$ as a function of $T$ for three different $q$, ($q=1$, $1.05$, $1.1$) as well as two different $\mathcal{U}$ ($\mathcal{U_{P}}$, $\mathcal{U_{L}}$). First of all, the finite-temperature QCD transition is not a real phase transition, but a crossover as Ref.~\cite{aoki2006order} shows, and it is independent of the $q$ and $\mathcal{U}$. Secondly, we find that the influence of $q$ on the chiral transition and deconfinement transition is consistent. That is to say, as $q$ increases, the transition occurs early, even though there will be obvious quantitative differences for two different $\mathcal{U}$. The same conclusion also appears in the non-extensive NJL model~\cite{Eur.Phys.J.A2016} and the non-extensive linear sigma model~\cite{shen2017chiral}.
In order to better study the influence of parameter $q$ on the crossover transition, we introduce the susceptibility, which is defined as
\begin{eqnarray}
\chi_{\sigma}=\frac{\partial\sigma}{\partial T}, \ \ \ \ \ \chi_{\Phi}=\frac{\partial\Phi}{\partial T}.
\end{eqnarray}
The peak position corresponds to the pseudo-critical temperature $T_{\chi}$ of the chiral transition and $T_{d}$ of the deconfinement transition, respectively. The results for two different $\mathcal{U}$ are presented in Tables~\ref{tb3},~\ref{tb4}.

From Tables~\ref{tb3},~\ref{tb4} we can clearly see that as $q$ increases, both $T_{\chi}$ and $T_{d}$ decrease. Specifically, as $q$ increases from 1 to 1.1, the pseudo-critical temperatures $T_{\chi}$ and $T_{d}$ decrease by approximately $6\%-8.2\%$ and $9.4\%-11.3\%$, respectively. Besides, we find $T_{\chi}>T_{d}$ and the same result can be seen in Ref~\cite{PhysRevD.81.074013}.
At last, we define a new susceptibility
\begin{figure}
\includegraphics[width=0.47\textwidth]{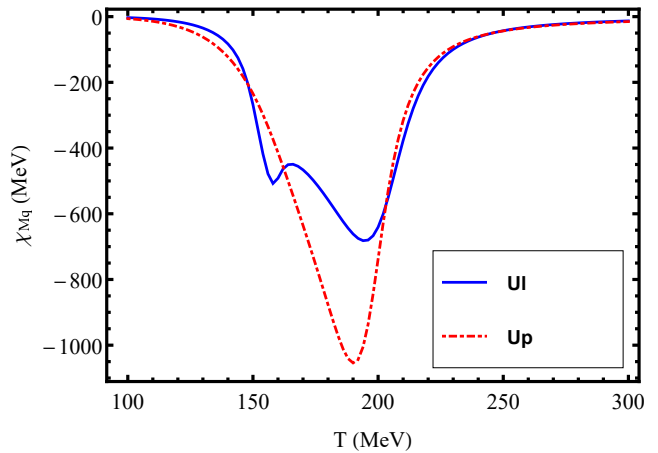}
\caption{The susceptibility $\chi_{Mq}$ as a function of $T$ at $\mu=0$ for two different Polyakov-loop potentials $\mathcal{U}$.}
\label{kaiq}
\end{figure}
\begin{eqnarray}
\chi_{xq}=\frac{\partial X}{\partial q},
\end{eqnarray}
to describe the response of $X$ to the parameter $q$. It can be seen from Fig.~\ref{kaiq} that for $M$, the maximum value of $|\chi_{Mq}|$ appears near the pseudo-critical temperature $T_{\chi}$. That is to say, at $T_{\chi}$ , $M$ has the largest response to $q$. However, it should be pointed out that in the case of the effective potential $\mathcal{U_{L}}$, the situation is slightly different. That is, $|\chi_{Mq}|$ will have maximum values near the pseudo-critical temperatures $T_{d}$ and $T_{\chi}$, respectively. Regarding $\chi_{\Phi q}$, it shows a maximum value near the pseudo-critical temperature $T_{d}$ for both effective potentials.
As for the response of other thermodynamic quantities and transport coefficients to $q$, we will discuss in detail in Sec. III.
\section{qcd thermodynamic quantities and transport coefficients}\label{bigtwo}
In this section, we mainly study the influence of parameter $q$ on the thermodynamic quantities and transport coefficients within q-PNJL model.
The reason why we are interested in thermodynamic quantities and transport coefficients is because they are not only sensitive to phase transition but also they can offer important information to other fields, like hydrodynamical models of the QGP, cosmological models of the early universe and models of massive objects in astrophysics as we emphasized above.
\subsection{QCD thermodynamic quantities}
\begin{figure}
\includegraphics[width=0.47\textwidth]{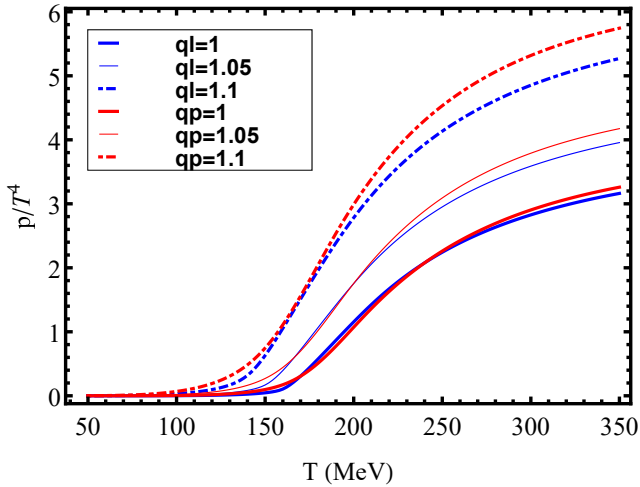}
\caption{The dimensionless pressure $p/T^{4}$ as a function of $T$ at $\mu=0$ for two different Polyakov-loop potentials $\mathcal{U}$ and three parameters $q$.}
\label{Fig:PT4}
\end{figure}
\begin{figure}
\includegraphics[width=0.47\textwidth]{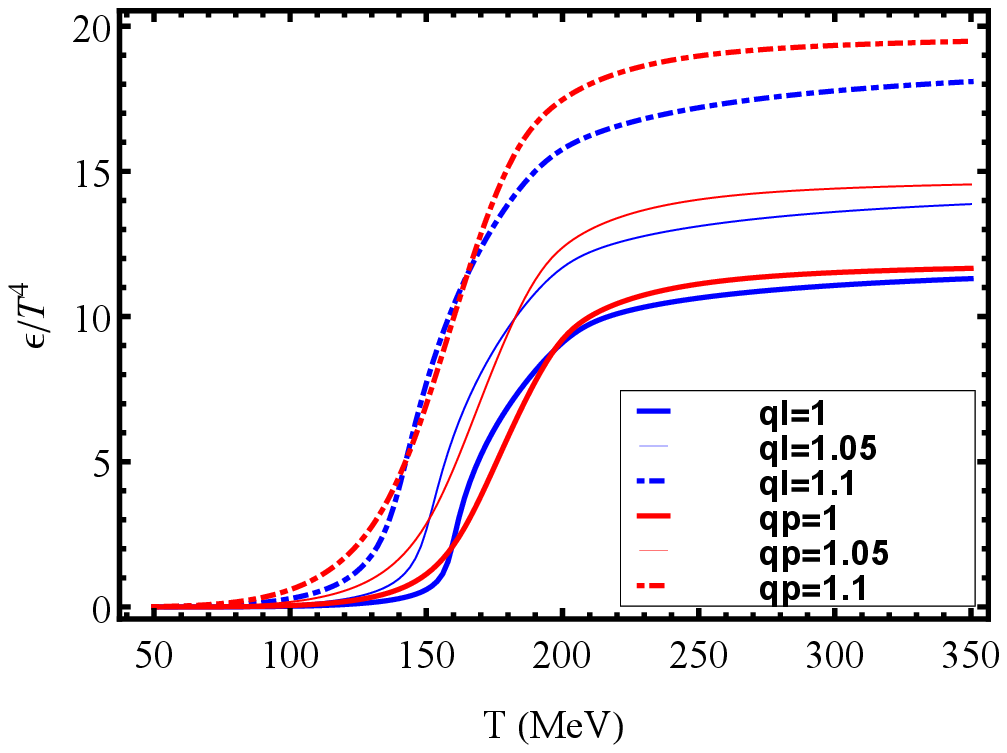}
\caption{The dimensionless energy density $\epsilon/T^{4}$ as a function of $T$ at $\mu=0$ for two different Polyakov-loop potentials $\mathcal{U}$ and three parameters $q$.}
\label{Fig:ET4}
\end{figure}
\begin{figure}
\includegraphics[width=0.47\textwidth]{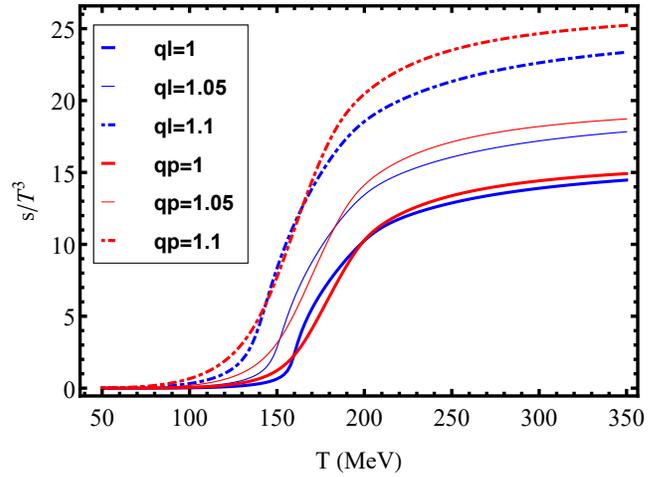}
\caption{The dimensionless entropy density $s/T^{3}$ as a function of $T$ at $\mu=0$ for two different Polyakov-loop potentials $\mathcal{U}$ and three parameters $q$.}
\label{Fig:ST3}
\end{figure}

All the thermodynamic information of a system is contained in the grand canonical potential which is given by $\Omega$ in Eq.~(\ref{omega00}) evaluated at the mean-field extent.
The pressure is
\begin{eqnarray}
p_{q}(T)=-\Omega_{q}(T),
\end{eqnarray}
with the vacuum normalization $p_{q}(0)=0$.
The entropy density $s_{q}$ and energy density $\epsilon_{q}$ are defined as follows
\begin{eqnarray}\label{sq}
s_{q}(T)=-\frac{\partial\Omega_{q}(T)}{\partial T}, \ \ \ \ \ \epsilon_{q}=-p_{q}(T)+Ts_{q}(T).
\end{eqnarray}

In the SB limit, the QCD pressure for $N_{c}^{2}-1$ massless gluons and $N_{f}$ massless quarks  is given by~\cite{PhysRevD.76.074023}
\begin{eqnarray}\label{SB}
\frac{p_{SB}}{T^{4}}&=&(N_{c}^{2}-1)\frac{\pi^{2}}{45}\nonumber\\
&&+N_{c}N_{f}[\frac{7\pi^{2}}{180}+\frac{1}{6}(\frac{\mu}{T})^{2}+\frac{1}{12\pi^{2}}(\frac{\mu}{T})^{4}],
\end{eqnarray}
where the first term is the gluonic contribution and the second term is the quark's contribution. For effective potentials $\mathcal{U_{P}}$, $\mathcal{U_{L}}$, at zero quark chemical potential and $N_{c}=3$, $N_{f}=2$, we have $p_{SB}/T^{4}\simeq4.06$.
Correspondingly,
\begin{eqnarray}
\epsilon_{SB}/T^{4}=3p_{SB}/T^{4}\simeq12.17,
\end{eqnarray}
and
\begin{eqnarray}
s_{SB}/T^{3}=p_{SB}/T^{4}+\epsilon_{SB}/T^{4}\simeq16.23.
\end{eqnarray}
\begin{figure}
\includegraphics[width=0.47\textwidth]{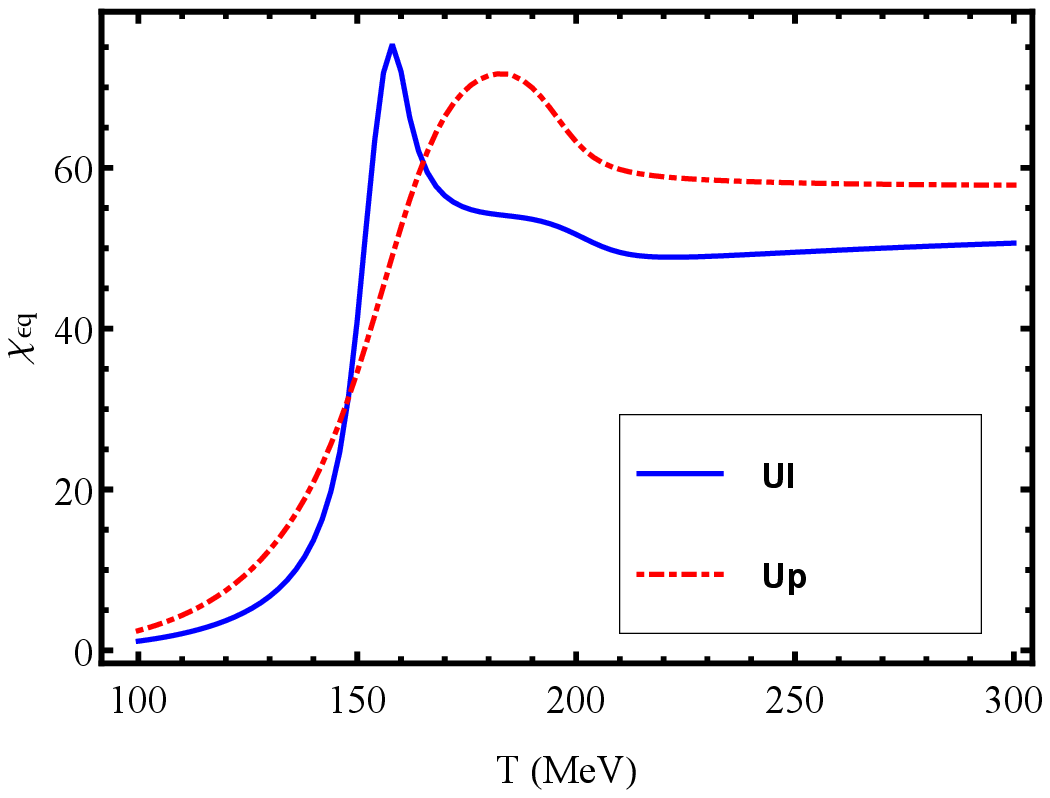}
\caption{The susceptibility $\chi_{\epsilon q}$ as a function of $T$ at $\mu=0$ for two different Polyakov-loop potentials $\mathcal{U}$.}
\label{energyT4kaiq}
\end{figure}
\begin{figure}
\includegraphics[width=0.47\textwidth]{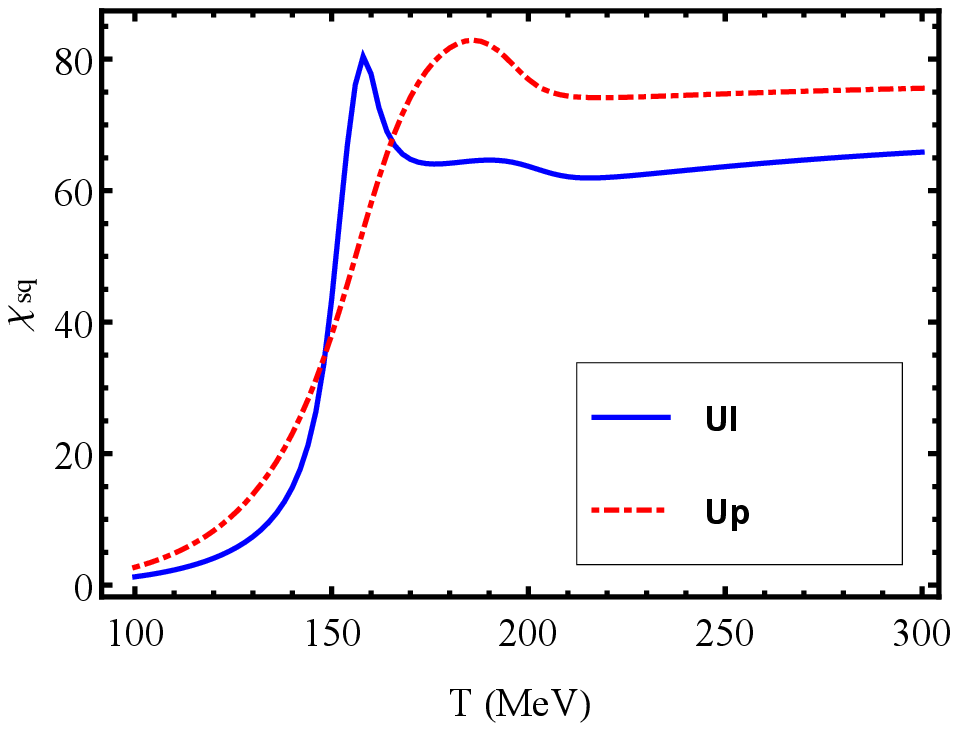}
\caption{The susceptibility $\chi_{sq}$ as a function of $T$ at $\mu=0$ for two different Polyakov-loop potentials $\mathcal{U}$.}
\label{kaisT3-q}
\end{figure}
\begin{figure}
\includegraphics[width=0.47\textwidth]{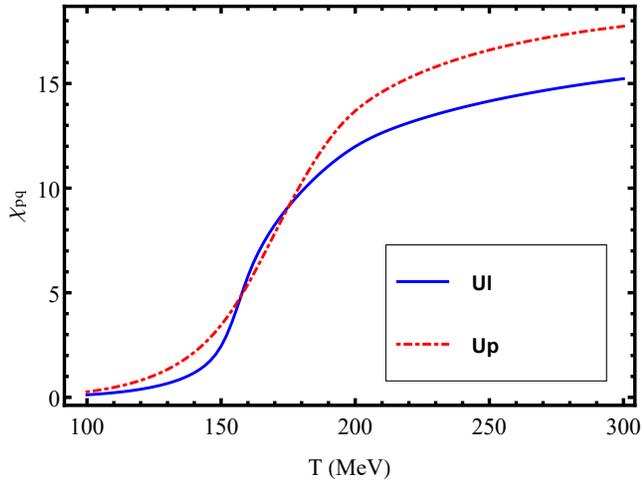}
\caption{The susceptibility $\chi_{pq}$ as a function of $T$ at $\mu=0$ for two different Polyakov-loop potentials $\mathcal{U}$.}
\label{PT4kaiq}
\end{figure}

From Figs.~\ref{Fig:PT4},~\ref{Fig:ET4},~\ref{Fig:ST3} we can see that when $q=1$, $p/T^{4}$, $\epsilon/T^{4}$ and $s/T^{3}$ all tend to their SB limit. However, as $q$ increases, they increase rapidly until they exceed their corresponding SB limits.
Taking $\epsilon/T^{4}$ as an example, for the Polyakov-loop potential $\mathcal{U_{L}}$ and the temperature is fixed at $350\ \mathrm{MeV}$. When $q=1$, the value of $\epsilon/T^{4}$ is 11.3, very close to the SB limit 12.18. But when $q=1.1$, the value of $\epsilon/T^{4}$ is 18.1, which is increased by 60\%. For $p/T^{4}$ and $s/T^{3}$, this value is 67\% and 61\%, respectively.
This is the result we expect because in the q-PNJL model, we use Tsallis statistics instead of BG statistics, and $q-1$ describes exactly the deviation from BG statistics. If $q-1$ is larger, the deviation from the SB limit is larger. That is to say, when a system is described by Tsallis statistics, then its high temperature limit is not the SB limit but the q-dependent Tsallis limit. This can be understood from the fact that the thermodynamic potential density function $\Omega_{q}\neq\Omega$ even when $T\rightarrow\infty$. That is to say, for the high temperature limit, the non-extensive effect still exists. This is different from the finite-size effect. Consider a cube system with a length of $L$, as $T$ increases, the thermal de Broglie wavelength decreases, and the effective size of the system becomes larger. Therefore, when $TL\rightarrow\infty$, we can still consider it as an ideal gas system and the related physical quantities tend to their SB limits. This means that the finite-size effect does not change the SB limit~\cite{PhysRevD.87.054009,Grunfeld2018,mogliacci2018geometrically}.
Furthermore, from Figs.~\ref{energyT4kaiq},~\ref{kaisT3-q}, we find that the response patterns of $\epsilon/T^{4}$ and $s/T^{3}$ for $q$ are almost the same. They reach a maximum near the pseudo-critical temperature $T_{c}$, then decrease and tend to be stable.
About $p/T^{4}$, from Eq.~(\ref{sq}), it is known that $\chi_{pq}=\chi_{sq}-\chi_{\epsilon q}$. The result is shown in Fig.~\ref{PT4kaiq}, in which its response increases with temperature and gradually stabilizes.

For relativistic heavy-ion collisions, the speed of sound is an important quantity, and its square at constant entropy is defined by
\begin{eqnarray}
c_{sq}^{2}=\frac{\partial p_{q}}{\partial\epsilon_{q}}\vert_{s_{q}}=\frac{\partial p_{q}}{\partial T}\vert_{V}/\frac{\partial \epsilon_{q}}{\partial T}\vert_{V}=\frac{s_{q}}{c_{vq}},
\end{eqnarray}
where $c_{vq}$ denotes the specific heat at constant volume, defined as
\begin{figure}
\includegraphics[width=0.47\textwidth]{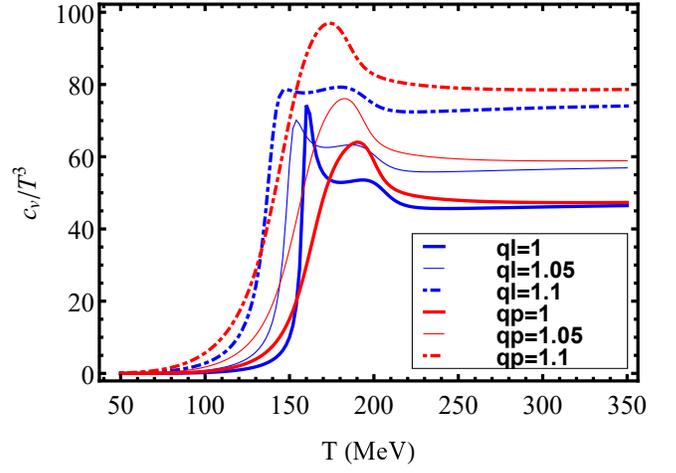}
\caption{The dimensionless specific heat $c_{v}/T^{3}$ as a function of $T$ at $\mu=0$ for two different potentials $\mathcal{U}$ and three parameters $q$.}
\label{Fig:CVT3}
\end{figure}
\begin{figure}
\includegraphics[width=0.47\textwidth]{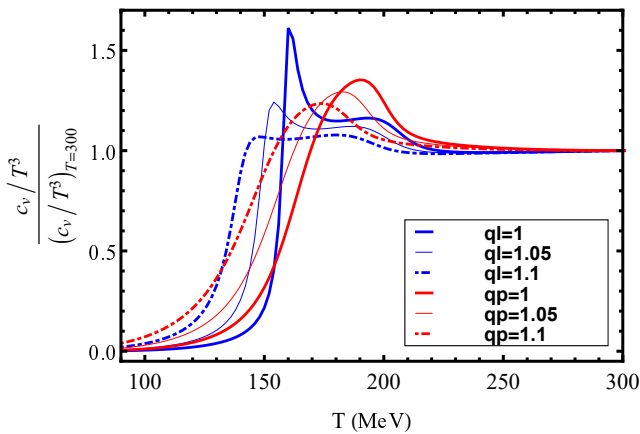}
\caption{The normalized dimensionless specific heat $c_{v}/T^{3}$ as a function of $T$ at $\mu=0$ for two different potentials $\mathcal{U}$ and three parameters $q$. The normalisation temperature is $T=300\ \mathrm{MeV}$.}
\label{Fig:NCVT3}
\end{figure}
\begin{figure}
\includegraphics[width=0.47\textwidth]{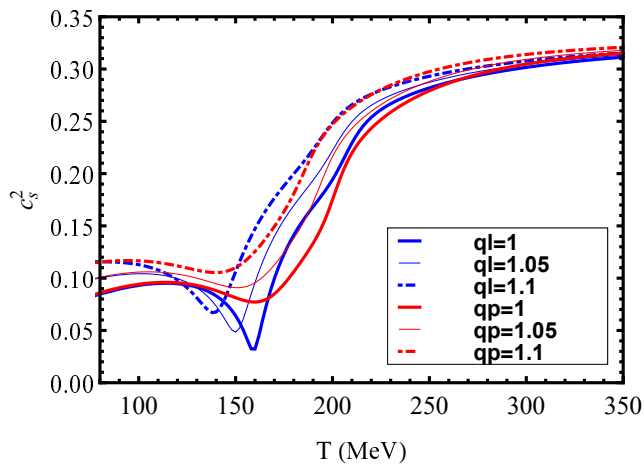}
\caption{The square of the speed of sound $c_{s}^{2}$ as a function of $T$ at $\mu=0$ for two different Polyakov-loop potentials $\mathcal{U}$ and three parameters $q$.}
\label{Fig:CS2}
\end{figure}
\begin{figure}
\includegraphics[width=0.47\textwidth]{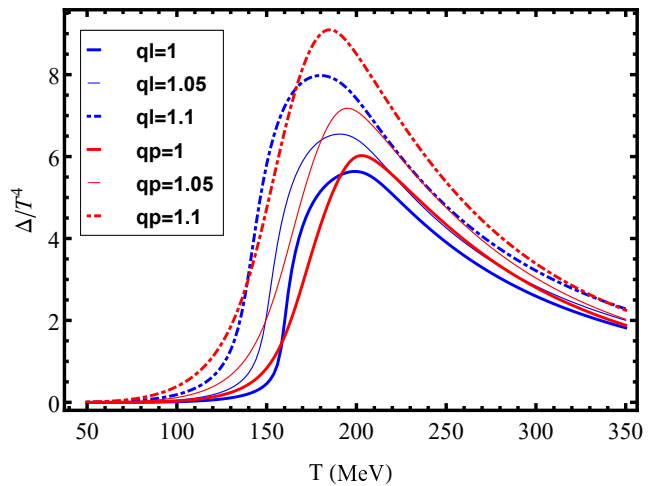}
\caption{The dimensionless interaction measure $\Delta/T^{4}$ as a function of $T$ at $\mu=0$ for two different Polyakov-loop potentials $\mathcal{U}$ and three parameters $q$.}
\label{Fig:deltaT4}
\end{figure}
\begin{eqnarray}
c_{vq}=\frac{\partial\epsilon_{q}}{\partial T}\mid_{V}=-T\frac{\partial^{2}\Omega_{q}}{\partial T^{2}}\mid_{V}.
\end{eqnarray}

From Fig.~\ref{Fig:CVT3}, like $p/T^{4}$, $\epsilon/T^{4}$ and $s/T^{3}$, when $q=1$, $c_{v}/T^{3}$ tends to the usual SB limit (48.69 for $\mathcal{U_{P}}$ and $\mathcal{U_{L}}$) with increasing temperature. However, as $q$ increases, the high temperature limit value also increases. In order to better show the effect of the non-extensivity parameter $q$ on $c_{v}/T^{3}$, it could be interesting to normalise all lines with respect to their high temperature limit, as shown in Fig.~\ref{Fig:NCVT3}. For $\mathcal{U_{P}}$, at $q=1$, the normalized $c_{v}/T^{3}$ starts to rise with increasing temperature $T$, then reaches the maximum near the pseudo-critical temperature $T_{\chi}$ of chiral transition, and eventually tends to the SB limit.
But for $\mathcal{U_{L}}$, there are two peaks. The first corresponds to the pseudo-critical temperature $T_{d}$ of deconfinement transition, and the second, although not obvious, corresponds to the pseudo-critical temperature $T_{\chi}$ of chiral transition as Refs.~\cite{Zhang_2018,Grunfeld2018} show.
Regarding the dependence of the normalized $c_{v}/T^{3}$ on $q$, the height and the pseudo-critical temperature $T_{c}$ of its peak decreases as $q$ increases for two Polyakov-loop potentials $\mathcal{U}$. Especially for $\mathcal{U_{L}}$, this peak flattens as $q$ increases so that it is difficult to show the pseudo-critical temperature $T_{c}$ of the crossover transition. That is to say, as $q$ increases, the critical behavior of $c_{v}/T^{3}$ is smoothed out and this phenomenon also appears in the $c_{s}^{2}$, which we will discuss next.

The behavior of $c_{s}^{2}$ is shown in Fig.~\ref{Fig:CS2}. For two different $\mathcal{U}$ and $q=1$, near the pseudo-critical temperature $T_{c}$ it has a dip and then approaches the ideal gas value of $1/3$ at high enough temperatures. The same conclusion appears in other models, such as the NJL model and the Polyakov-Quark-Meson (PQM) model~\cite{PhysRevD.81.074013,PhysRevC.88.045204}. Besides, as the $q$ increases, we can see that the dip is gradually disappearing. This is similar to Ref.~\cite{Khuntia2016} where it calculated the speed of sound as a function of temperature for different q-values for a hadron resonance gas and found that when $q$ is larger than $1.13$, all criticality disappears. Interestingly, similar phenomena have appeared in the study of finite-size effects. Refs.~\cite{PhysRevD.87.054009,Grunfeld2018,PhysRevD.97.116020} indicate that as the size decreases, the critical behavior of $c_{s}^{2}$ also gradually or even completely disappears. It is reasonable to agree with each other because the finite-size effect is part of the non-extensive effect. In addition, it is worth noting that due to a surprising cancellation, the high temperature limit of $c_{s}^{2}$ is still its SB limit $1/3$, independent of $\mathcal{U}$. Taking $\mathcal{U_{L}}$ as an example, we find that at high temperature $T=350\ \mathrm{MeV}$, when $q$ increases to 1.1, $s/T^{3}$ and $c_{v}/T^{3}$ increase by 61.4\% and 59.4\%, respectively. As the temperature further increases, the growth rates of $s/T^{3}$ and $c_{v}/T^{3}$ tend to be the same. Therefore, the high temperature limit of $c_{s}^{2}=\frac{s}{T^{3}}/\frac{c_{v}}{T^{3}}$ is not affected by non-extensive effects.

Another quantity that is related to $c_{s}^{2}$ is the interaction measure $\Delta_{q}(T)$ which measures the deviation from the equation of state of an ideal gas $\epsilon=3p$ due to interactions and/or finite quark masses, defined as
\begin{eqnarray}
\Delta_{q}(T)=\epsilon_{q}(T)-3p_{q}(T),
\end{eqnarray}
and it is related to the $c_{s}^{2}$ via
\begin{eqnarray}\label{cs2}
\frac{\Delta_{q}(T)}{\epsilon_{q}(T)}=\frac{\epsilon_{q}(T)-3p_{q}(T)}{\epsilon_{q}(T)}\simeq1-3c_{sq}^{2}.
\end{eqnarray}
The reason for doing this approximation comes from Refs~\cite{PhysRevD.73.114007,PhysRevD.81.074013}, from which we can see that $c_{s}^{2}$ and $p/\epsilon$ are in good agreement at both low temperature ($<T_{c}$) and high temperature ($>2.5T_{c}$). At the intermediate temperature ($T_{c}\sim 2.5T_{c}$), $c_{s}^{2}$ is slightly larger than $p/\epsilon$.
Therefore, it can be known from Eq.~(\ref{cs2}) that near the pseudo-critical temperature $T_{c}$, the minimum value of $c_{s}^{2}$ will cause the maximum value of $\Delta_{q}(T)$. At the high temperature limit, $c_{s}^{2}$ tends to $1/3$ and $\Delta_{q}(T)$ tends to zero.
From Fig.~\ref{Fig:deltaT4} we can clearly see that it has a peak near the pseudo-critical temperature $T_{c}$ and then tends to zero. And the peak moves with $q$ towards a lower temperature, which is similar to $c_{v}/T^{3}$ and independent of $\mathcal{U}$.
\subsection{Transport coefficients}
\begin{figure}
\includegraphics[width=0.47\textwidth]{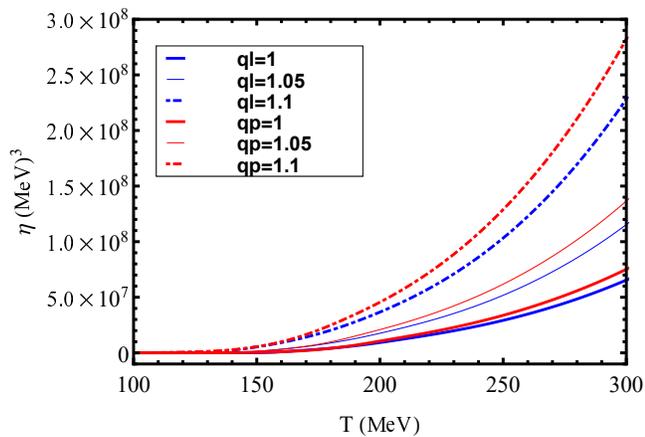}
\caption{The shear viscosity $\eta$ as a function of $T$ at $\mu=0$ for two different Polyakov-loop potentials $\mathcal{U}$ and three parameters $q$.}
\label{Fig:eta-q}
\end{figure}
\begin{figure}
\includegraphics[width=0.47\textwidth]{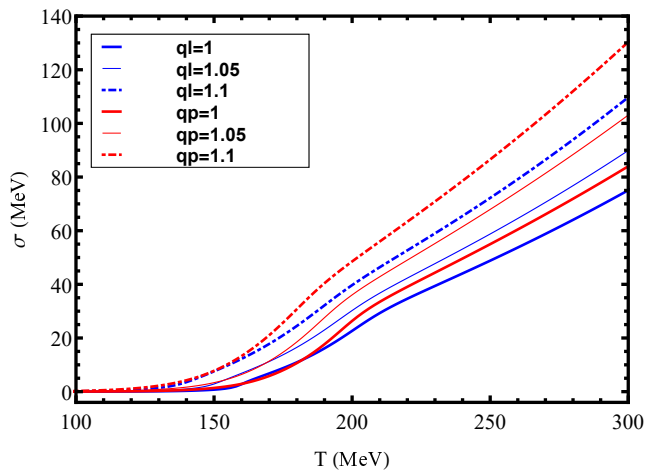}
\caption{The electrical conductivity $\sigma$ as a function of $T$ at $\mu=0$ for two different Polyakov-loop potentials $\mathcal{U}$ and three parameters $q$.}
\label{Fig:sigma-q}
\end{figure}
\begin{figure}
\includegraphics[width=0.47\textwidth]{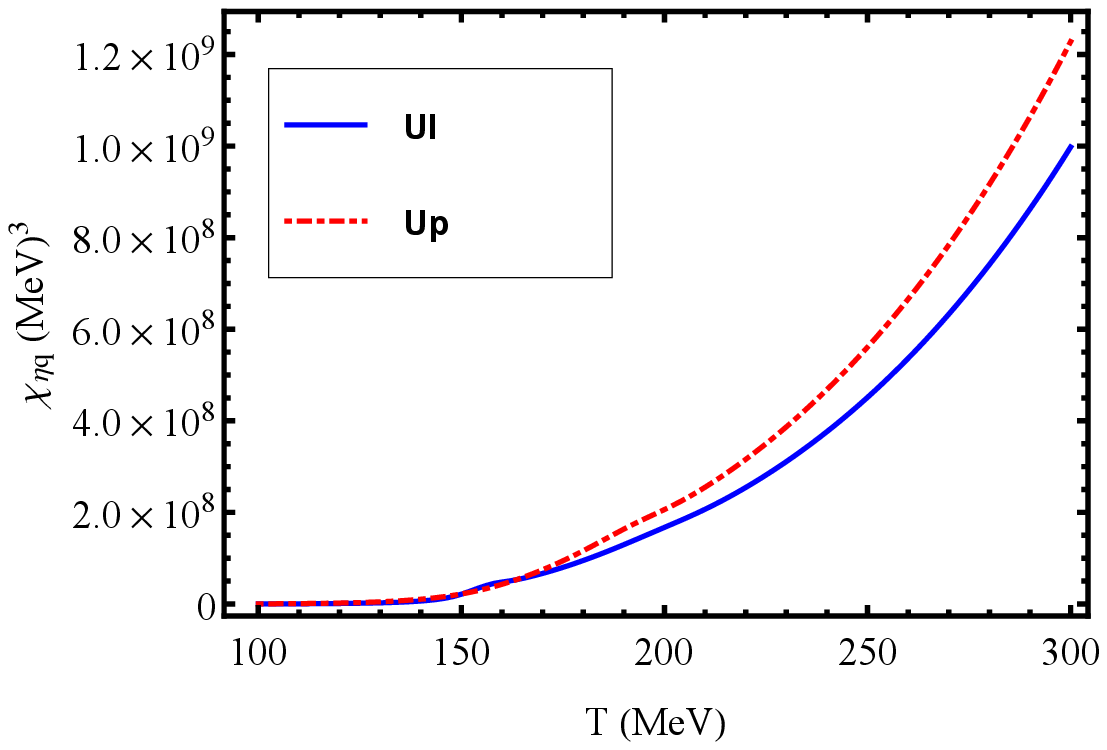}
\caption{The susceptibility $\chi_{\eta q}$ as a function of $T$ at $\mu=0$ for two different Polyakov-loop potentials $\mathcal{U}$.}
\label{Fig:kaieta-q}
\end{figure}
\begin{figure}
\includegraphics[width=0.47\textwidth]{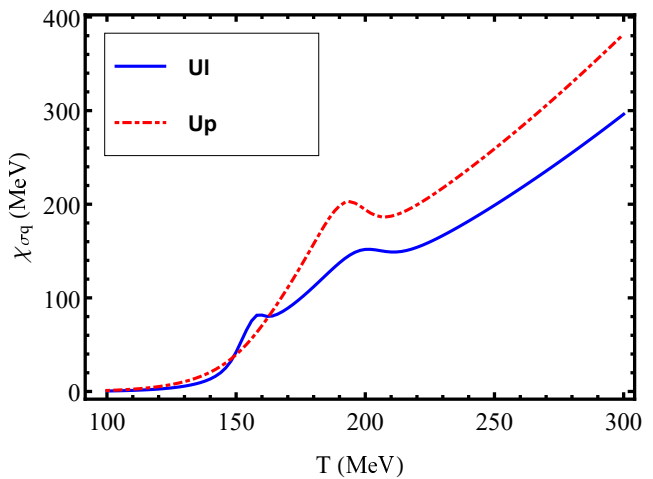}
\caption{The susceptibility $\chi_{\sigma q}$ as a function of $T$ at $\mu=0$ for two different Polyakov-loop potentials $\mathcal{U}$.}
\label{Fig:kaisigma-q}
\end{figure}
In this subsection we are mainly concerned with the influence of nonextensivity parameter $q$ on the transport coefficients, such as shear viscosity $\eta$, electrical conductivity $\sigma$ and bulk viscosity $\zeta$. Based on linear $\sigma$ model~\cite{PhysRevC.83.014906}, NJL model~\cite{PhysRevC.93.045205,SASAKI201062,PhysRevC.88.068201,Lang2014,Lang2015,PhysRevD.94.094002}, PQM model~\cite{PhysRevD.97.014005}, PNJL model~\cite{PhysRevD.91.054005} and the Parton-Hadron-String Dynamics (PHSD) transport approach~\cite{PhysRevLett.110.182301}, we get a gross summary about the temperature dependence of these transport coefficients. $\eta(T)$ and $\sigma(T)$ decrease with temperature increase in the hadronic phase, while they increase with temperature increase in the QGP phase, and show a minimum at the transition temperature. For certain materials, like helium, nitrogen and water, this temperature dependence of $\eta$ has been experimentally confirmed~\cite{PhysRevLett.97.152303}. While $\zeta(T)$ follows an opposite trend, which shows a maximum at the transition temperature~\cite{PhysRevC.83.014906,SASAKI201062,PhysRevD.94.094002,Xiao_2014}.
The mathematical expressions of transport coefficients
calculated from relaxation time approximation (RTA) in
the kinetic theory approach and calculated from the one-loop diagram approximation in the quasi-particle Kubo approach are equivalent, as follows~\cite{PhysRevD.97.116020,PhysRevD.99.014004}
\begin{widetext}
\begin{eqnarray}
\eta&=&\frac{2N_{C}N_{f}}{15T}\int\frac{{\rm d}^3\vec{p}}{(2\pi)^3\Gamma}(\frac{\vec{p}^{2}}{E_{p}})^{2}[n_{q}(1-n_{q})+\bar{n}_{q}(1-\bar{n}_{q})],\\
\sigma&=&(\frac{2N_{c}}{3T})(\frac{5e^{2}}{9})\int\frac{{\rm d}^3\vec{p}}{(2\pi)^3\Gamma}(\frac{\vec{p}}{E_{p}})^{2}[n_{q}(1-n_{q})+\bar{n}_{q}(1-\bar{n}_{q})],\\
\zeta&=&\frac{2N_{c}N_{f}}{T}\int\frac{{\rm d}^3\vec{p}}{(2\pi)^3\Gamma}\frac{1}{E_{p}^{2}}\{(\frac{1}{3}-c_{sq}^{2})\vec{p}^{2}-c_{sq}^{2}M^{2}+c_{s}^{2}MT\frac{dM}{dT}\}^{2}[n_{q}(1-n_{q})+\bar{n}_{q}(1-\bar{n}_{q})].
\end{eqnarray}
\end{widetext}

\begin{figure}
\includegraphics[width=0.47\textwidth]{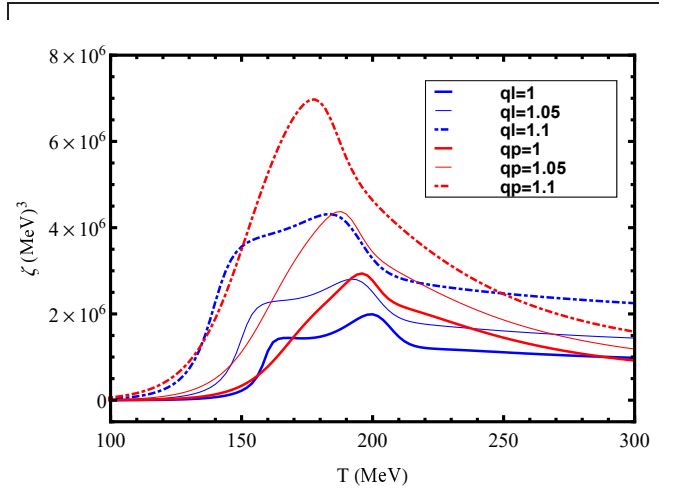}
\caption{The bulk viscosity $\zeta$ as a function of $T$ at $\mu=0$ for two different Polyakov-loop potentials $\mathcal{U}$ and three parameters $q$.}
\label{Fig:ksi-q}
\end{figure}

It should be noted that $n_{q}$ and $\bar{n}_{q}$ are not the usual Fermi-Dirac distribution but the q-version of the Fermi-Dirac distribution, defined by Eqs.~(\ref{nq}),~(\ref{nbarq}). In addition, all our discussions are based on a constant value of relaxation time $\tau=1/\Gamma=1\ \mathrm{fm}$.
The changes of $\eta$ and $\sigma$ with $T$ and three parameters $q$ are shown in Figs.~\ref{Fig:eta-q},~\ref{Fig:sigma-q}. First of all, we find that $\eta$ and $\sigma$ rise monotonically with temperature increase as Refs.~\cite{PhysRevD.97.116020,PhysRevD.99.014004} show, and they increase as $q$ increases. More specifically, for $\eta$ , its response to $q$ increases monotonically with temperature, as shown in Fig.~\ref{Fig:kaieta-q}. For $\sigma$, its response to $q$ appears an extreme value near the pseudo-critical temperature and then rises monotonically, as shown in Fig.~\ref{Fig:kaisigma-q}.
Regarding $\zeta$, from Fig.~\ref{Fig:ksi-q} we find that the situation will be different for different $\mathcal{U}$. For example, in the effective potential $\mathcal{U_{P}}$, $\zeta$ increases with temperature and has a significant maximum near the pseudo-critical temperature $T_{c}$, and then tends to zero. However, in the effective potential $\mathcal{U_{L}}$, $\zeta$ will have two maximum values as the temperature rises. The first maximum value is not so obvious, corresponding to the vicinity of the pseudo-critical temperature $T_{d}$ of deconfinement transition, and the second maximum value corresponds to the vicinity of the pseudo-critical temperature $T_{\chi}$ of chiral transition. A similar double-peak structure also appears in the NJL model~\cite{PhysRevC.93.045205} and PNJL model~\cite{PhysRevD.97.116020}. Moreover, Ref.~\cite{PhysRevD.97.116020} indicates that the double-peak structure disappears when the size is reduced to $2\ \mathrm{fm}$. This is similar to our results, that is, as $q$ increases, the two peaks begin to merge into a broad one.
And this phenomenon is very similar to $c_{v}/T^{3}$ because the transport coefficient $\zeta$ is related to $c_{s}^{2}$ and therefore also related to $c_{v}/T^{3}$.
Besides, its response to $q$ has a maximum near the pseudo-critical temperature $T_{c}$.

\begin{figure}
\includegraphics[width=0.47\textwidth]{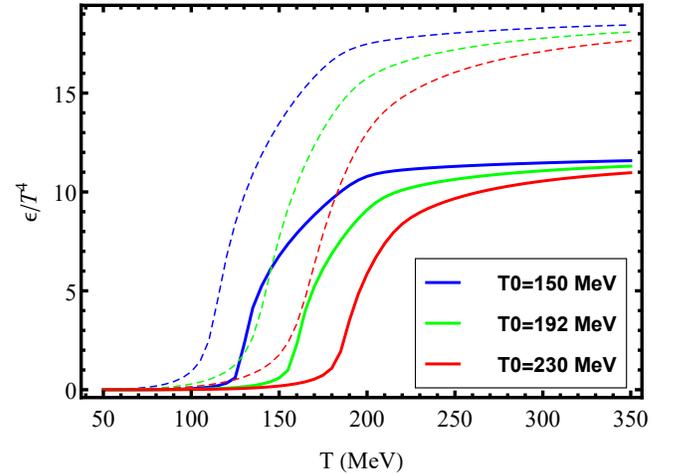}
\caption{The dimensionless energy density $\epsilon/T^{4}$ as a function of $T$ at $\mu=0$ for $\mathcal{U_{L}}$, three $T_{0}$ and two $q$. The thick line represents $q=1$ and the dashed line represents $q=1.1$.}
\label{Fig:compare}
\end{figure}

\begin{figure}
\includegraphics[width=0.47\textwidth]{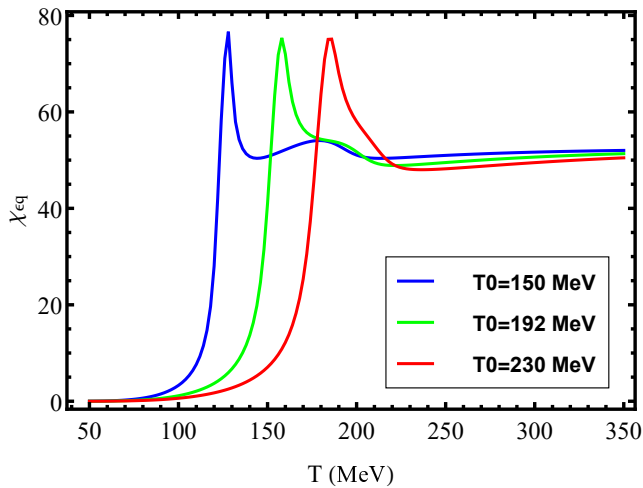}
\caption{The susceptibility $\chi_{\epsilon q}$ as a function of $T$ at $\mu=0$ for $\mathcal{U_{L}}$, three $T_{0}$.}
\label{Fig:compare1}
\end{figure}

Finally, the value of $T_{0}$ cannot be completely determined, so we consider $T_{0}$ as a free parameter to test the $T_{0}$-dependence of our results. We find that $T_{0}$ has only a quantitative effect on the results, mainly manifested in that the pseudo-critical temperature $T_{\chi}$ and $T_{d}$ decreases as $T_{0}$ decreases, as pointed out in Ref.~\cite{PhysRevD.81.074013}. However, our results are qualitatively independent of $T_{0}$. Taking $\epsilon/T^{4}$ as an example, as shown in Figs.~\ref{Fig:compare},~\ref{Fig:compare1}, we find that its change with $q$ is not affected by $T_{0}$.
\section{Summary and Conclusion}\label{summary}
In this paper, combined with the Tsallis statistics and the PNJL model, we investigated the sensitivity of phase transitions, thermodynamic quantities, and transport coefficients to deviations from usual BG statistics. It was found that the chiral and deconfinement transition are still a crossover at finite temperature and zero quark chemical potential, independent of the non-extensivity parameter $q$. However, their corresponding pseudo-critical temperatures $T_{\chi}$ and $T_{d}$ decrease as $q$ increases.
Regarding the influence of the parameter $q$ on the thermodynamic quantities. On the one hand, we found that for $\frac{\epsilon}{T^{4}}$, $\frac{p}{T^{4}}$ and $\frac{s}{T^{3}}$, their high temperature limit is no longer the SB limit but the q-related Tsallis limit. But for $c_{s}^{2}$, due to a surprising cancellation, its high temperature limit is not affected by the parameter $q$. It should be noted that the finite-size effect does not change the SB limit. On the other hand, we found that as the $q$ increases, the criticality of $c_{v}/T^{3}$ and $c_{s}^{2}$ will gradually or even completely disappear. This is consistent with the finite-size effect, where the same phenomenon occurs as the size decreases.

Under a constant value of relaxation time $\tau=1/\Gamma=1\ \mathrm{fm}$, we calculated the transport coefficient as a function of temperature and found that $\eta$ and $\sigma$ rise monotonically with temperature increase, while $\zeta$ shows a maximum near the pseudo-critical temperature $T_{c}$. About the influence of the parameter $q$ on them, we found that $\eta$ and $\sigma$ increase as $q$ increases at any fixed temperature, while $\zeta$ changes with $q$, similar to $c_{v}/T^{3}$. We also introduced a new susceptibility in order to study the response of thermodynamic quantities and transport coefficients to $q$ in more detail. We found that their response patterns to $q$ are different. For example, $\epsilon/T^{4}$, $s/T^{3}$, $\Delta/T^{4}$ and $c_{s}^{2}$ have the highest response to $q$ near the pseudo-critical temperature $T_{c}$, while $\sigma$ shows a maximum and a minimum values near the $T_{c}$ and then rises monotonically with temperature increase. For $p/T^{4}$, its response increases with temperature increase and gradually stabilizes. Besides, it should be noted that the double-peak structure appearing in $c_{v}/T^{3}$ and $\zeta$ is unique to Polyakov-loop potential $\mathcal{U_{L}}$. And as $q$ increases, it gradually disappears. Interestingly, in the study of finite-size effects, the double-peak structure also disappears as the size decreases.

As a first step, we are only concerned with the zero quark chemical potential and finite temperature region and $q>1$. Next, we will study the whole $q$-dependence of the equation of state at finite temperature and finite quark chemical potential to further study its influence on the properties of protoneutron stars~\cite{Lavagno2011}. Furthermore, determining the existence and location of the critical end point (CEP) in QCD phase transition has been one of the main goals of relativistic heavy-ion collision experiments. For this purpose, the second phase of the beam energy scan at RHIC will be performed between 2019 and 2021~\cite{LUO201675}. Therefore, the impact of non-extensive effect on the location of CEP is a question worthy of further study~\cite{shen2017chiral}. Moreover, Ref.~\cite{PhysRevD.87.076004} proposes an improved Polyakov-loop potential due to the backreaction of the quarks. Thus, it is also an interesting question to use the improved Polyakov-loop potential to study the non-extensive effect to compare with the usual Polyakov-loop potential.
Finally, in order to qualitatively understand the influence of the parameter $q$ on the transport coefficients, we have taken constant value of relaxation time in this present work. However, involved calculations of relaxation time at finite temperature and finite quark chemical potential in Tsallis statistics, incorporating different interaction channels might lead us to more realistic scenario. These issues are our future research directions.
\acknowledgments
We sincerely thank the referee for the detailed, comprehensive and enlightening suggestions. This has greatly helped the improvement of our article.
\bibliography{PRD}
\end{document}